%% file: arxiv.tex
\documentclass{article}
\usepackage{arxiv}
\usepackage[utf8]{inputenc} % allow utf-8 input
\usepackage[T1]{fontenc}    % use 8-bit T1 fonts
\usepackage{url}            % simple URL typesetting
\usepackage{booktabs}       % professional-quality tables
\usepackage{amsfonts}       % blackboard math symbols
\usepackage{nicefrac}       % compact symbols for 1/2, etc.
\usepackage{microtype}      % microtypography
\usepackage{lipsum}
\usepackage{amssymb}
\usepackage{latexsym}

\usepackage{url}
\usepackage{xcolor}

\usepackage[pagebackref=true,breaklinks=true,letterpaper=true,colorlinks,bookmarks=false]{hyperref}

\definecolor{newcolor}{rgb}{.8,.349,.1}

\input{math_commands.tex}

\usepackage{amsfonts}       % blackboard math symbols
\usepackage{nicefrac}       % compact symbols for 1/2, etc.
\usepackage{microtype}      % microtypography
\usepackage{booktabs}
\usepackage{algpseudocode}
\usepackage{algorithm}
\usepackage{mathtools}
\usepackage{color}
\usepackage{multirow,graphicx,paralist,bigdelim}
\usepackage{colortbl}
\usepackage{tabularx}
\usepackage{comment}
\usepackage{epstopdf}
\usepackage{epsfig}
\usepackage{graphicx}
\usepackage{amsmath}
\usepackage{amssymb}
\usepackage{pdfrender}
\usepackage{wrapfig}
\usepackage{tikz}
\usepackage{enumitem}
\usepackage{comment}
\usepackage{tablefootnote}

\date{}
\title{Fibro-CoSANet: Pulmonary Fibrosis Prognosis Prediction using a Convolutional Self Attention Network}

\author{
 Zabir Al Nazi \\
  Brainekt AI Lab\\
  Dhaka, Bangladesh\\
  \texttt{zabiralnazi@yahoo.com} \\
  
  \And 
 Fazla Rabbi Mashrur\\
  Department of Biomedical Engineering\\
  Khulna University of Engineering \& Technology\\
  Khulna, Bangladesh \\
  \texttt{rabbi.mashrur@gmail.com} \\
  
  \And
 Md Amirul Islam \\
  Department of CS, Ryerson University\\
  Vector Institute for AI\\
  Toronto, Canada\\
  \texttt{amirul@cs.ryerson.ca} \\
  
   \And
  Shumit Saha\thanks{Corresponding Author; Email: shumit.saha@mail.utoronto.ca
} \\
  University Health Network (UHN)\\
  University of Toronto\\
  Toronto, Canada\\
  \texttt{shumit.saha@mail.utoronto.ca} \\
}

\begin{document}

\maketitle

\begin{abstract}
\input abstract.tex
\end{abstract}

% keywords can be removed

\input introduction.tex

\input approach.tex
\input experiment.tex
\input discussion.tex

\bibliographystyle{unsrt}  
\bibliography{ct}
\end{document}

%% file: math_commands.tex
\usepackage{amsmath,amsfonts,bm}

\def\eqref#1{equation~\ref{#1}}
\def\1{\bm{1}}

\DeclareMathAlphabet{\mathsfit}{\encodingdefault}{\sfdefault}{m}{sl}
\SetMathAlphabet{\mathsfit}{bold}{\encodingdefault}{\sfdefault}{bx}{n}

%% file: abstract.tex
Idiopathic pulmonary fibrosis (IPF) is a restrictive interstitial lung disease that causes lung function decline by lung tissue scarring. Although lung function decline is assessed by the forced vital capacity (FVC), determining the accurate progression of IPF remains a challenge. To address this challenge, we proposed Fibro-CoSANet, a novel end-to-end multi-modal learning based approach, to predict the FVC decline. Fibro-CoSANet utilized CT images and demographic information in convolutional neural network frameworks with a stacked attention layer. Extensive experiments on the OSIC Pulmonary Fibrosis Progression Dataset demonstrated the superiority of our proposed Fibro-CoSANet by achieving new state-of-the-art modified Laplace Log-Likelihood score of -6.68. This network may benefit research areas concerned with designing networks to improve the prognostic accuracy of IPF. The source-code for Fibro-CoSANet is available at: \url{https://github.com/zabir-nabil/Fibro-CoSANet}.

%can aid to improve the prognostic accuracy of IPF and better manage the disorder. The source-code for Fibro-CoSANet is available at: \url{https://github.com/zabir-nabil/Fibro-CoSANet}.

%which outperformed the existing state-of-the-art methods

%% file: introduction.tex
\section{Introduction}
Idiopathic pulmonary fibrosis (IPF) is a \textcolor{black}{chronic}  lung disease which is caused by forming scar tissue within the lungs \cite{spagnolo_idiopathic_2015}. IPF leads to a gradual, irreversible deterioration of lung function by replacing the healthy lung tissues with scar tissue over time. IPF can potentially lead to rapid deterioration from long-term stability, which results in complete pulmonary dysfunction \cite{raghu_diagnosis_2018}. Due to the high variability in deterioration speed, \textcolor{black}{management of pulmonary fibrosis relies on the decline in the lung function progression}. Therefore, \textcolor{black}{ an accurate estimation of the lung function progression decline would lead to better management of IPF}.

\textcolor{black}{The current guideline for IPF diagnosis follows several procedures, such as surgical or transbronchial lung biopsy \cite{raghu_diagnosis_2018}. After the diagnosis, physicians often assess the decline of lung function by Force vital capacity (FVC) using spirometry tests to monitor the prognosis of IPF. FVC measures the total amount of air exhaled after breathing in as deeply as possible \cite{zappala2010marginal}. To assess the lung function, observing the FVC at intervals of six to twelve months is recommended~\cite{raghu_diagnosis_2018}.} \textcolor{black} {While the FVC provides a general understanding of the prognosis of the IPF \cite{flaherty_idiopathic_2006}, there are no widely used techniques to estimate the IPF progression. As such, due to the heterogeneous course of IPF, imaging modalities may provide valuable information regarding IPF prognosis.} 

Computed tomography (CT) images of the chest can be effectively used to assess the lung function decline from pulmonary fibrosis as the CT scans contain several visual signs essential for assessment by radiologists. Shi et al. \cite{shi_prediction_2019} developed a voxel-wise radio-logical model using high-resolution CT scans and achieved $82.1\%$ accuracy in predicting the progression of IPF. Furthermore, Salisbury et al. \cite{salisbury_idiopathic_2016} utilized CT scans of IPF patients to predict the survival and FVC decline for 12 months with a significant correlation value of 0.6 between visual and predicted measurement. These studies have demonstrated the effectiveness of utilizing CT imaging as an important modality to predict the progression of pulmonary fibrosis. However, precisely predicting the progression of IPF from CT images remains challenging due to the high variability. 

The recent advancements of artificial intelligence (e.g., convolutional neural networks (CNNs)~\cite{he2016deep}) and the Kaggle: OSIC Pulmonary Fibrosis Progression Challenge \cite{osic_kaggle} have significantly inspired to develop CT image based machine learning systems to obtain computer-aided clinical decision for IPF prognosis. 
In particular, Wong et al. \cite{wong_fibrosis-net_2021} recently proposed Fibrosis-Net based on deep CNNs for predicting pulmonary fibrosis progression from chest CT images. Fibrosis-Net utilized the chest CT scans of a patient along with spirometry measurement and clinical metadata to predict the FVC of a patient at a specific time-point in the future~\cite{wong_fibrosis-net_2021}. While the existing CNNs based approaches have a higher capacity to predict pulmonary fibrosis progression from chest CT images, we strongly believe there is still room for improvement in terms of overall correctness. In this work, we argue that extracted convolutional features from chest CT scans along with patient's clinical or demographic features are not discriminative enough to correctly predict the FVC of a patient in cases where the network requires to focus on a specific region of the lung. To address this issue, we proposed a simple and efficient end-to-end multi-modal network, termed as Fibro-CoSANet, which utilized both the chest CT scan images and demographic information, such as sex, age, smoking history to predict the FVC of a patient at a specific time-point. Our proposed Fibro-CoSANet used a convolutional self-attention network that extracted features from a randomly selected CT image which are merged with the normalized demographics features. The merged features were then passed through a one-layer perceptron to obtain the predicted FVC. While the Fibrosis-Net~\cite{wong_fibrosis-net_2021} utilized the multiple CT slices to generate convolutional features, we introduced an efficient formulation of the IPF prognosis task where we randomly selected a single CT image from multiple scans to extract convolutional features. However, we used the approximated lung volume information from all the available scans as a shallow feature which was merged with the convolutional features. In addition, we predicted the \textit{slope} of FVC based on a \textit{linear prior assumption} to reduce the computational overhead, while Fibrosis-Net~\cite{wong_fibrosis-net_2021} used an elastic net to obtain the local FVCs. 

We summarize our main contributions as follows:
\begin{itemize}

\item To the best of our knowledge, this is the first study that proposed a simple and efficient end-to-end multi-modal based convolutional self-attention network to predict the progress of IPF by utilizing the deeper CT and shallower demographic features. 

\item We introduced an intuitive and efficient way to apply a stacked self-attention layer on top of extracted convolutional CT features for further refinement and the advantages of this module are demonstrated with extensive experiments.

\item We further introduced a unique formulation for FVC measurement of a patient where the goal of the proposed network was to predict the \textit{slope} of the FVC trend. 

\item We showed, through extensive quantitative experiments under different settings, that our proposed Fibro-CoSANet achieved lower modified Laplace
Log-Likelihood score than existing works on the publicly available Kaggle: OSIC Pulmonary Fibrosis Progression dataset.

\end{itemize}

%% file: approach.tex
\vspace{-5pt}
\section{Data Preparation}
In this section, we discuss the number of preprocessing steps conducted to prepare the training inputs and labels.  
We used the recently introduced Kaggle: OSIC Pulmonary Fibrosis Progression Challenge Benchmark Dataset which consists of CT scans, FVC measurements, and associated demographic features, such as age, sex, smoking status\cite{osic_kaggle}. As the main goal of our method was to predict the \textit{slope} of the FVC trend of a patient, we first calculated the initial \textit{slope} of FVC values using singular value decomposition (SVD) \cite{golub_singular_1971} which were used as pseudo-labels in our proposed model (see Fig. \ref{fig:data_pipeline} (3)). Then, we estimated the lung volume from the CT scans (Fig. \ref{fig:data_pipeline} (1)) followed by the extraction and normalization of demographics, such as age, sex, and smoking status (Fig. \ref{fig:data_pipeline} (2)). Note that, we trained our proposed Fibro-CoSANet using a random CT image, estimated lung volume, age, sex, and smoking status.

\vspace{-5pt}

\subsection{Forced Vital Capacity (FVC) Formulation}  \label{slope}
We introduced a unique formulation to predict the \textit{slope} of the FVCs by using the calculated initial \textit{slope} of FVC values as ground-truth. 
First, we pre-processed the CT scans, $C$ $\in$ \{$c_i, c_{i+1},...., c_n$\}, where $n$ refers to the number of patients. For each CT scan, $c_i$ $\in$ \{$s_1, s_2,...., s_{m_{i}}$\}, we randomly selected a slice, $s_k$, from $c_i$ for extracting features, where $m_{i}$ is the number of slices in $c_i$ and $k$ is the index of the selected slice. The selected slice, $s_k$, was fed to a self-attention driven CNN model to extract pulmonary CT features. In parallel, we generated the demographics and volumetric feature sets, $H$, where $H$ $\in$ \{$h_1, h_2,...., h_n$\}. Finally, these two sets of features were concatenated to predict the \textit{slope} of FVC, $Z$ $\in$ \{$z_i, z_{i+1},..., z_n$\}, based on a linear priori assumption. Each patient data had FVC values, $z_i$ $\in$ \{$v_{w_1}, v_{w_2},... ,v_{w_{t_{i}}}$\}, where $t_{i}$ and $w$ refers to the number of FVC values and representation of the corresponding weeks, respectively. We can formalize the FVC value of the $i^{th}$ patient in $j^{th}$ week as follows:

%%%%%%%%%%%%%%%%%%%%%%%%%%%
\begin{figure} [t]
\centering

    {\includegraphics[width=0.7\textwidth]{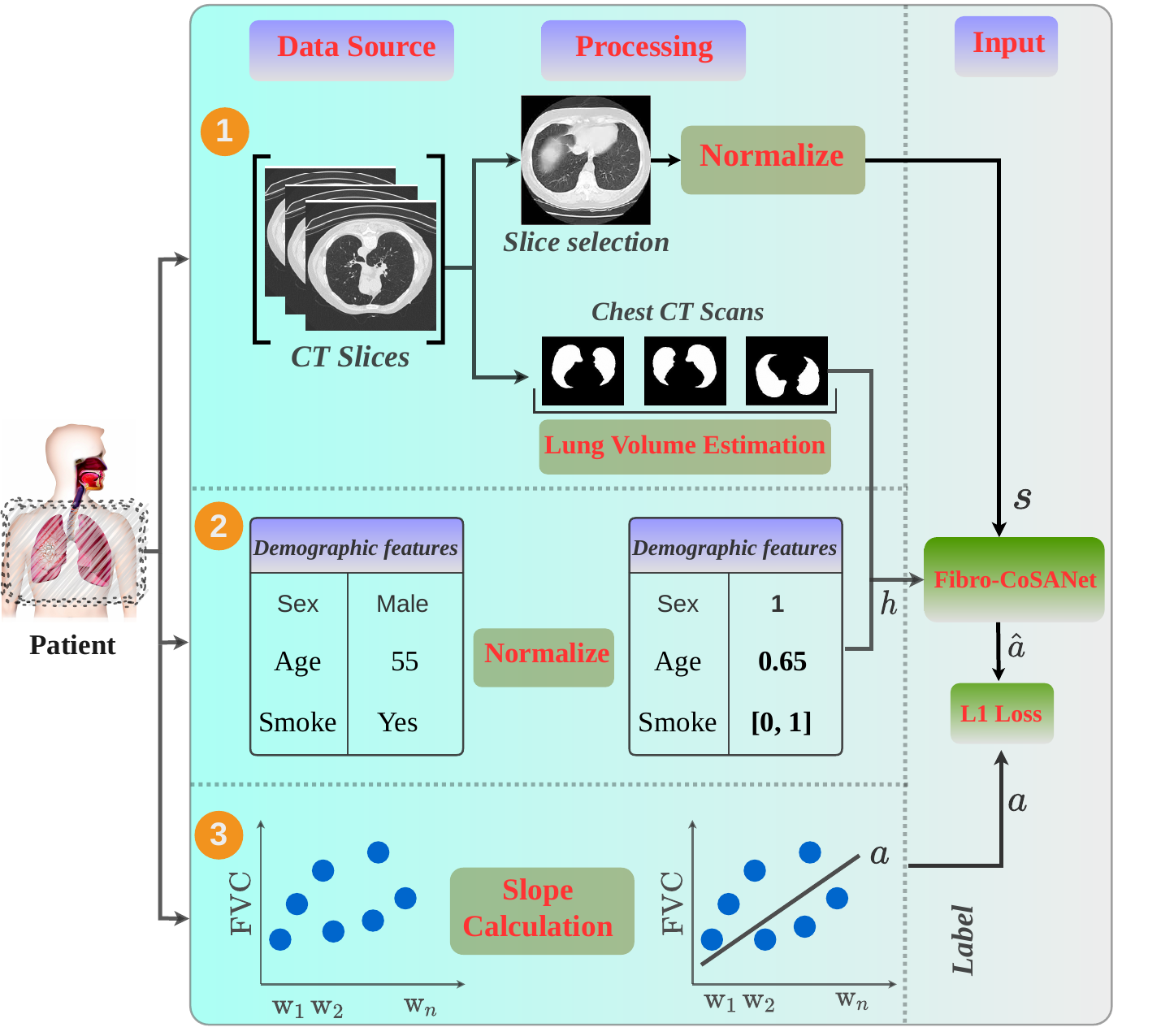}}
    \caption{Illustration of the process of dataset preparation (\textcircled{1}, \textcircled{2}) and ground-truth generation (\textcircled{3}) for training Fibro-CoSANet. We randomly selected a CT slice and passed through the Fibro-CoSANet. We further normalized the demographic features and estimated the lung volume to fuse with the convolutional CT features. We calculated the \textit{slope} of FVC using SVD least squares method explained in Sec. \ref{svd_exp} which is used as ground-truth to measure the correctness of the predicted FVC \textit{slope}.
    }
     \label{fig:data_pipeline}
    \vspace{-10pt}
\end{figure}
%%%%%%%%%%%%%%%%%%%%%%%%%%%%%
\begin{equation} \label{eq:eq1}
    V_{w_{j}} = a_i \times{w_j}+ V_{b_{i}}
\end{equation}
\label{svd_exp}
where, $V_{b_{i}}$ is the base FVC and $a_i$ is the \textit{slope} of the $i^{th}$ patient. We can further extend Eq.~\ref{eq:eq1} by expanding FVC along the week, upto $j$, as follows: 

\begin{equation} \label{eq:eq2}
    w_{1}  a_i+V_{b_i}=V_{w_{1}}, \hspace{0.2cm} w_{2}  a_i+V_{b i}=V_{w_{2}}, ......, {w}_{j}  a_i+v_{b i}=V_{w_{j}}
\end{equation}
For ease of presentation, we vectorized Eq.~\ref{eq:eq2} as follows:

\begin{equation} \label{eq:eq4}
A x = b \hspace{0.2cm} \text{where} \hspace{0.2cm}
A=\begin{bmatrix}
w_{1} & 1 \\
w_{2} & 1 \\
\vdots & \vdots \\
w_{j} & 1
\end{bmatrix}, \hspace{0.2cm} 
x=\begin{bmatrix}
a_i & V_{b_{i}}
\end{bmatrix}, \hspace{0.2cm}
b=\begin{bmatrix}
V_{w_{1}}  \\
V_{w_{2}}  \\
\vdots \\
V_{w_{j}} 
\end{bmatrix} 
\end{equation}

\noindent Next, we decomposed the matrix $A_{j \times 2}$ into singular value form as: 

\begin{equation} \label{eq:eq5}
    A_{j \times 2} = U \Sigma V^{\top}
\end{equation}

where $U$ and  $\Sigma$ refer to $j \times j$ orthogonal matrix and $j \times 2$ diagonal matrix, respectively. $V^{\top}$ is a $2 \times 2$ orthogonal matrix. We replaced $A$ in Eq.~\ref{eq:eq4} with Eq.~\ref{eq:eq5} to achieve our desired least square solution \cite{golub_singular_1971}.
The replacement operations can be formalized and solved using singular value decomposition \cite{golub_singular_1971}  as follows:
\begin{equation}
\begin{aligned}
\label{svd}
Ax &= b, \hspace{0.3cm}
\tilde x &= A^{+}b_{j \times 2} 
\end{aligned}
\end{equation}

\noindent where $A^{+}$ is Moore–Penrose inverse of the matrix $A_{j \times 2}$ and $ \tilde x$ minimizes our desired least square solution, $\|A \tilde{x}-b\|_{2}$. Thus, we calculated the \textit{slope}, $a_i$, for a patient $i$, and used it as a pseudo ground-truth slope to train our network.

\vspace{-5pt}
\label{ct_pre}
\subsection{CT Pre-processing and Lung Volume Extraction}

Contrary to natural images, CT scans consist of inconsistent and high-dimensional redundant information \cite{park2020annotated} which is computationally expensive to process and can result in poor performance. Therefore, to achieve a better signal-to-noise ratio, it is imperative to pre-process the CT scan images before feeding them to the CNN model. We applied the following pre-processing steps to resolve the issues.\\
\noindent \textbf{Slice Selection.} Each patients' CT scan, $c_{i}$, contains many CT slices which represent the depth information of the lung. To reduce the computational complexity, we selected one slice per CT scan by the following operations: (i) We first truncated the first and last $15\%$ of the slices as these slices contain minimal volume information. (ii) Then, we \textit{randomly} select one CT slice, $s_{k}$, from the middle to feed into the CNN. 
    
\noindent \textbf{Resizing and Normalizing CT Slices.} We resized the randomly chosen CT slice based on the input specification $(512 \times 512)$ of the backbone CNN model. Further, to mitigate the inconsistency in tissue intensities across different scanners and improve convergence of the model, we normalized the pixel values using, $s' = (s - \lambda_{b})/ (\lambda_{a} - \lambda_{b})$, where $\lambda_{a} = 2048$ and $\lambda_{b} = 0$ for any CT slice, $s$. \\
\noindent \textbf{Lung Volume Estimation from the CT Slices.} \label{vol_calc}
Along with demographics, we calculated the lung volume from CT images for each patient. The main motivation of using the lung volume estimation was to incorporate the approximated volume in the feature set, as we didn't include all the CT slices to extract the volumetric features due to the computational complexity. Given a CT slice, $s$, we applied the \textit{watershed} algorithm  \cite{beucher_use_1979} to extract a segmentation map, $p$, of the lung for slice, $s$. The generated map, $p$, is a binary map between $\{0, 1\}$, where 1 and 0 indicate if a pixel belongs to the lung or not. Then, we generated the segmented lung image by simply multiplying the binary segmentation map with the original CT image. The procedure of generating the lung volume, $v_{i}$, for $i^{th}$ patient can be formalized as follows: 

\begin{equation}
    v_{i} = \sum_{j=1}^{m_{i}} p_{j} \times \delta_{x} \times \delta_{y} \times \delta_{z}
\end{equation}
where $\delta_x$ and $\delta_y$ are the pixel spacing in $x$ and $y$ directions respectively, and $\delta_z$ denotes the thickness of the slice.
\vspace{-5pt}
\subsection{Extracting Demographic Features}

As IPF is associated with demographics, such as baseline FVC \cite{raghu_diagnosis_2018,ley_multidimensional_2012}, age \cite{garcia-sancho_familial_2011,ley_multidimensional_2012}, gender \cite{garcia-sancho_familial_2011}, smoking status \cite{garcia-sancho_familial_2011}, we take inspiration from~\cite{wong_fibrosis-net_2021} to incorporate these features along with CT image to improve the performance of our proposed method. 
We normalized the estimated lung volume, age, sex, and smoking status features using: $ x^{\prime}=\frac{x-\bar{x}}{\sigma}$ where, $x$ is the raw numeric feature, $\bar{x}$ is the arithmetic mean of $x$, and $\sigma$ is the standard deviation of $x$.

\begin{figure} [t]
\centering
    {\includegraphics[width=0.99\textwidth]{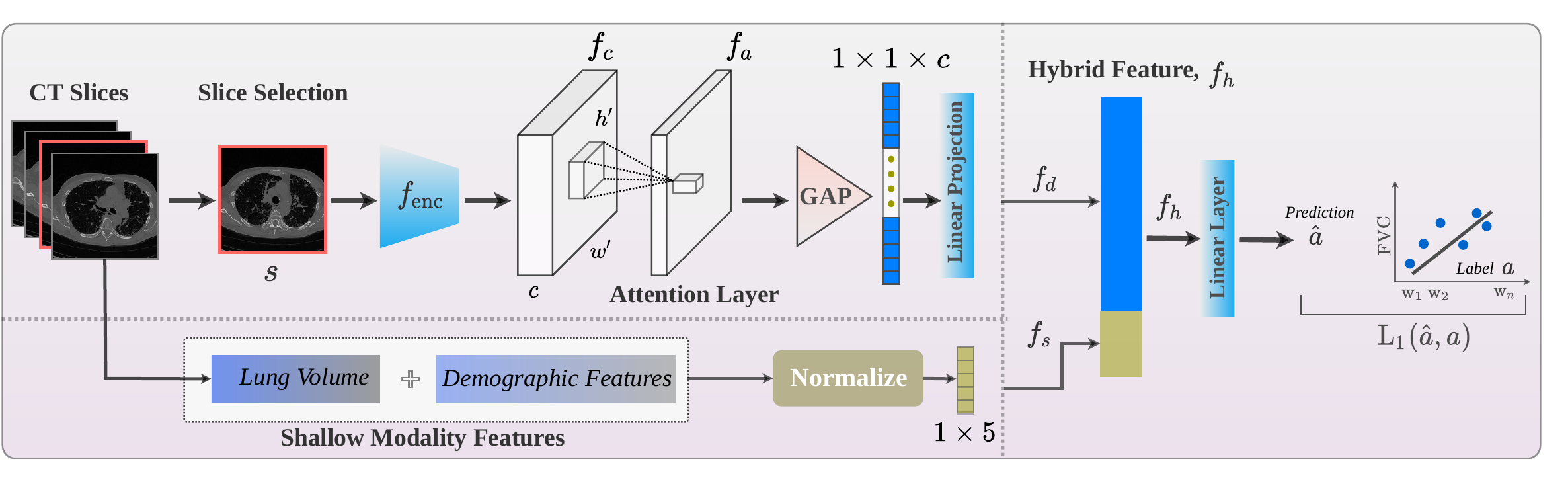}}
    \caption{Illustration of the dual-stream pipeline of our proposed Fibro-CoSANet. We passed the randomly selected CT image to a pre-trained CNNs, $f_{\text{enc}}$, (e.g., ResNet18~\cite{he_deep_2016}) which produced a CT feature map, $f_c$. Then, we integrated a stacked self-attention module on top of the last convolutional layer of CNN to allow the network to focus on specific regions. The resultant output feature map, $f_a$, from the stacked self-attention module is passed through the global average pooling to produce the final representation, $f_d$, of the CT image. 
    In parallel, we merged the demographics and estimated lung volume features, $f_a$ with $f_d$ which is further passed through a linear layer to predict the 
    \textit{slope} of FVC. We computed $L1$ loss between the predicted \textit{slope}, $\hat{a}$ and pseudo ground-truth \textit{slope}, $a$.}.
    \label{fig:workflow}
    \vspace{-5pt}
\end{figure}
\vspace{-5pt}
\section{Network Architecture of Fibro-CoSANet}

We proposed a novel multi-modal convolutional self-attention network, Fibro-CoSANet, to predict the \textit{slope} of FVC in an end-to-end manner. The overall pipeline of our proposed Fibro-CoSANet is illustrated in Fig.~\ref{fig:workflow}. Our proposed training framework consists of two key steps: (i) Extraction of the deep features from the normalized CT image using a CNN with self-attention module (Sec. \ref{convsa}), (ii) Fusing the deep features extracted from a CNN with shallow lung volume and demographic features followed by a fully-connected layer which predicts the \textit{slope} of the FVC (Sec. \ref{ref:fusion}).

\vspace{-5pt}
\subsection{Convolutional Self-Attention Network}
\label{convsa}

In this section, we present our proposed convolutional self-attention network for extracting features from CT scan images with the ultimate goal of predicting the \textit{slope} of FVC. Our proposed deep CT feature extractor network consists of two key components, (i) a CNN-based feature extractor network and (ii) a self-attention module which further refined the convolutional features extracted from the CNN. We first discuss the convolutional feature extractor network (Sec. \ref{deepcnn}) followed by the self-attention module (Sec. \ref{selfattention}).

\subsubsection{Deep CNN for CT Feature Extraction}
\label{deepcnn}

In recent years, CNNs have been widely adopted for processing medical images (e.g., CT scans) \cite{sarvamangalaconvolutional}. In general, CNN-based networks on medical imaging can be characterized as generic feature extractor networks which are termed encoder networks. The encoder network is simply a CNN that extracts features from a given input image. However, one downside of training CNNs is that it requires a huge amount of labeled training data to learn the millions of parameters involved in the network. This crucial issue limits the adoption of CNNs on medical image-based tasks as the majority of the datasets have a small volume of training data. To address this limitation, inspired by the existing works \cite{ sajja2019lung, wang2020classification}, we fine-tuned the feature extractor CNN on CT scan images rather than training from scratch with random initialization. Let $s\in \mathbb{R}^{h\times w \times 1}$ be the input CT scan image (where $h$, $w$ are the spatial dimensions). Given the input CT image, $s$, we adopted a CNN, $f_{\text{enc}}$, to extract a feature representation from the last convolutional layer of the CNN. Let $f_c\in \mathbb{R}^{h^\prime \times w^\prime \times c}$ be the extracted feature map which has smaller spatial dimensions than the original CT image, $s$. We used \textit{ResNet} \cite{he2016deep}, \textit{ResNeXt} \cite{Xie2016}, and \textit{EfficientNet}  \cite{tan2019efficientnet} based architectures to build encoder networks in our study. We formalize the key operations as follows:
% \cite{harsono2020lung, sajja2019lung, wang2020classification},

\begin{equation}
    f_c = f_{\text{enc}} (\mathbf{W}_a \ast s)
\end{equation}
where $\mathbf{W}_a$ denotes weights of the CNN model and $\ast$ denotes the convolutional operation. The extracted feature map, $f_c$, was fed to a self-attention module which further refines the feature representations before combining them with the demographics and lung volume features.

\subsubsection{Self-Attention Module}
\label{selfattention}

The extracted feature map, $f_c$, was likely to capture high-level semantics of the CT images; however, allowing the network to focus on the specific region of the CT feature map was important to accurately predict the progression of IPF. Since IPF can result in honeycomb cysts in the lungs \cite{gruden_ct_2016}, these regions in the CT image require more attention than others. To focus on these regions of interest, we took inspiration from the existing works~\cite{ramachandran_stand-alone_2019,zhang_self-attention_2019} and applied a self-attention module on top of the CNN feature extractor.

In the self attention module, we first rearranged the extracted convolutional feature map, $f_c$, resulting in a feature map, $f_c' \in \mathbb{R}^{c'  \times  N}$ where $c'$ is the number of channels and $N$ is the product of $h'$ and $w'$. Then we fed the feature map, $f_c'$, to the self-attention layer~\cite{zhang_self-attention_2019}, and obtained an attention map, $o_a \in \mathbb{R}^{c'  \times  N}$, with same dimension.

We further multiplied the output of the attention layer, $o_a$, by a scaling parameter, $\gamma$, and added with the input feature map, $f_c'$ to obatin the self-attention map, $f_a$. We formalized the operations as follows:
\begin{equation}
    \boldsymbol{f}_{a}=\gamma * \boldsymbol{o}_{a} + f_c'
\end{equation}

Note that $\gamma$ is a learnable scalar that is initialized from a uniform distribution in our work. The main advantage of learning $\gamma$ was that it enabled the network to first focus on the local neighborhood indicators since it was easier. Then it eventually tried to assign more weight to the non-local region. Thus the module learned simpler tasks first to improve convergence \cite{zhang_self-attention_2019}. Finally, $f_a$ was passed through an adaptive global average pooling operation (GAP), followed by a linear layer to obtain the deep CT feature set, $f_d$ (see Fig.~\ref{fig:workflow}). We considered $f_d$ as our final feature representation extracted from the CT scan image. % is worth noting that the refinement of the convolutional feature map by applying a self-attention layer was reported performing well in recent work \cite{ramachandran_stand-alone_2019}.

We augmented the CNN with the self-attention layer for realizing a richer effective receptive field and learning better feature representation as the recent work~\cite{ramachandran_stand-alone_2019} has shown the advantages of applying a self-attention layer on top of convolutional feature representation. Unlike the previous self-attention-based works~\cite{ramachandran_stand-alone_2019,zhang_self-attention_2019}, we extended the existing idea by stacking $l$ self-attention layers just before applying the GAP operation. In our implementation, $l$ is the stacking factor and we considered $l$ as a hyper-parameter. Note that, we placed the attention module between the last convolutional layer of CNNs and the pooling layer as convolutions were likely to better capture the low-level features while stand-alone attention layers may integrate global information by modeling long range pixel interactions~\cite{ramachandran_stand-alone_2019}. Furthermore, this placement reduced the computational complexity as the attention module was applied on a relatively low dimensional convolutional feature map.

\begin{figure}
\centering
    {\includegraphics[width=0.8\textwidth]{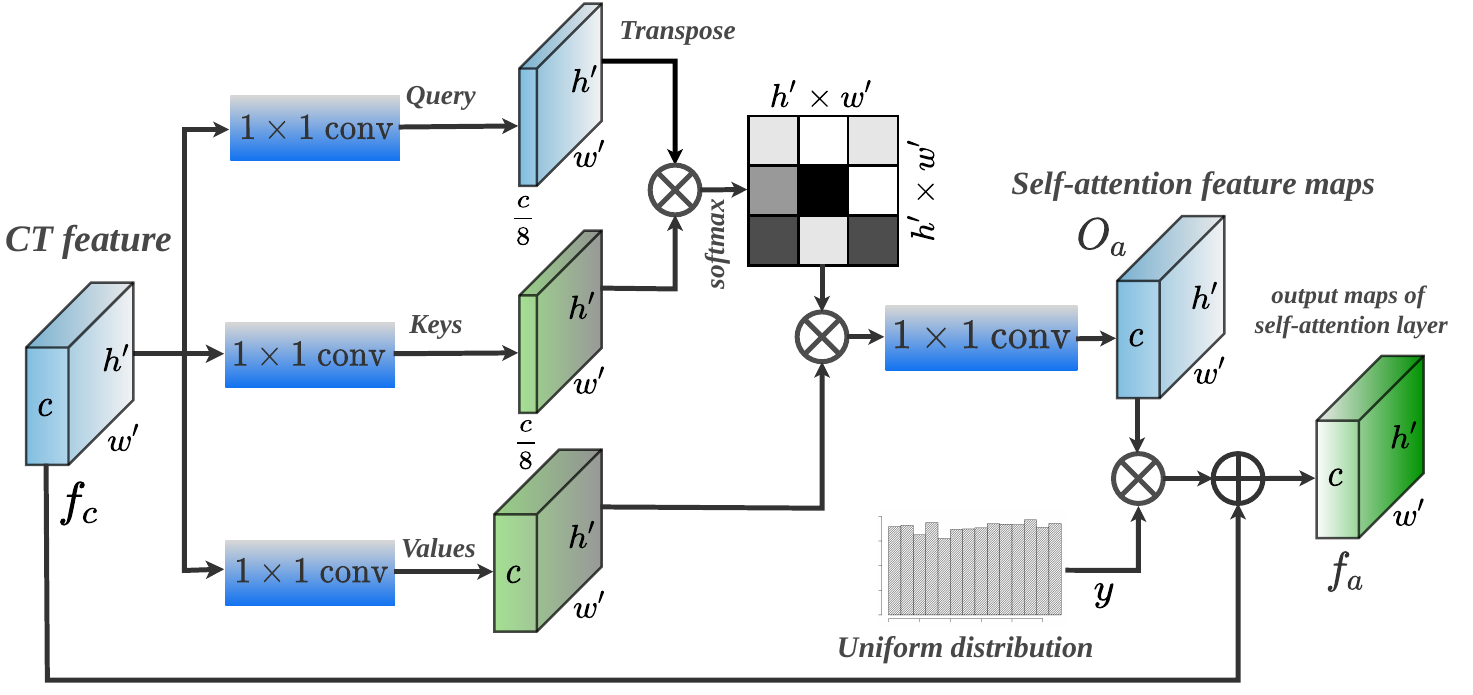}}
    %\vspace{-0.5cm}
    \caption{Overview of the proposed self-attention module for Fibro-CoSANet. Note that the stacking factor is one here.}
    \label{fig:attention}
    \vspace{-10pt}
\end{figure}

\vspace{-5pt}
\subsection{Hybrid Fusion of Convolutional and Shallow Modality Feature}
\label{ref:fusion}
Finally, we concatenated both the deep CNN features, $f_d$, extracted by the self-attention driven CNN backbone from the CT modality and the shallow modality features representation, $f_s$, to generate a hybrid multi-modal feature representation, $f_h$. The resultant feature representation, $f_h$, was passed to a fully connected layer to obtain the \textit{slope}, $a$, of FVC, which was used to predict the patient's progression curve. We computed FVC from the predicted \textit{slope}, $a$ along the timeline, $w$, as follows:
\begin{equation}
    \text{FVC}(w_j) = a \times w_j + \text{FVC}_{b}
\end{equation}

where $\text{FVC}_{b} $ is the baseline force vital capacity and $j$ is the week index. \footnote{Patient index $i$ is used with variables when specifying a particular subject, for general case, the index is omitted for simplicity.}
\label{var_conf}

%% file: experiment.tex
\vspace{-5pt}
\section{Experiments}
We evaluated the effectiveness of our proposed approach for predicting the progression of pulmonary fibrosis and demonstrated the efficacy of the method under different settings. First, we showed the superiority of our proposed multi-modal learning pipeline followed by a comparison with recent approaches. Then, we evaluated our approach to generating baseline results under different backbone and metric settings to show generalizability and consistency. We further conducted an ablation study to investigate the necessity of each component of our proposed approach. Finally, we provided a comparison between different variants of our approach in terms of computational complexity, inference time, and total memory. \\

\noindent \textbf{Dataset.} In this study, we used the publicly available Kaggle: OSIC Pulmonary Fibrosis Progression challenge benchmark dataset provided by Open Source Imaging Consortium (OSIC) \cite{osic_kaggle}. The dataset consists of chest CT scans and associated demographics about fibrosis diagnosed patients. It contains 176 unique patients with a total of 1576 demographic information (multiple from the same patients) collected from numerous follow-up visits over the course of approximately 1-2 years. The demographics include the patient's \textit{ID, weeks, FVC, percent, age, sex, and smoking status}. Note that the \textit{weeks} represent the relative number of weeks pre or post from the baseline CT scan for each patient and we determined the time series of the weeks of a specific patient based on the patient's ID. For each patient, CT scan images (varies between 10 -180) are provided in \textit{DICOM} format files that contained meta-data about the patients and the scan. We used $5$ fold cross-validation scheme to validate the best performance model. In the cross-validation setting, we carefully restrict to have no overlapping between the train and test splits of different subjects. The test set includes a baseline CT scan with \textit{only} the initial FVC measurement for each patient.

\noindent \textbf{Evaluation Metric.}  We used a modified Laplace Log-Likelihood ($LLL_{m}$) and root mean square error ($RMSE$) metrics to report the performance of our proposed model. We choose $LLL_{m}$ to evaluate a model's confidence in its decisions as it is designed to reflect both the accuracy and certainty of each prediction. For each true FVC measurement, we calculated the FVC and confidence measure as follows \cite{osic_kaggle}:

\begin{equation}
\begin{array}{c}
\sigma_{\text {clipped}}=\max (\sigma, 70) \\
\Delta=\min \left(\left|F V C_{\text {true}}-F V C_{\text {predicted}}\right|, 1000\right) \\
LLL_{m}= -\frac{\sqrt{2} \Delta}{\sigma_{\text {clipped}}}-\ln \left(\sqrt{2} \sigma_{\text {clipped}}\right)
\end{array}
\end{equation}

where $\sigma$ is the standard deviation and we threshold the error at 1000 ml to avoid the adverse penalty due to large errors. The confidence values were clipped at 70 ml to reflect the approximate measurement uncertainty in FVC. We calculated the final score by averaging the metric across all \textit{weeks}. Note that, the calculated value of the metric was always negative, and \textit{lower is better}. 

% RMSE

%%%%%%%%%%%%%%%%%%%%%%%%%%%%%%%%%%%%%%
\begin{table}[tb]
\label{multimodal}
\caption{Performance comparison between different modalities. Our proposed multi-modality based Fibro-CoSANet outperforms the single modalities (e.g., CT and Demographics).}
\scriptsize
\centering
\label{tab:modality}
\def\arraystretch{1.2}
\setlength\tabcolsep{3.9pt}
\resizebox{0.60\textwidth}{!}{
\begin{tabular}{c|c|c}

\specialrule{1pt}{.5pt}{.5pt}\
Mode & $LLL_m \downarrow$  & $RMSE \downarrow$  \\ \specialrule{1pt}{.5pt}{.5pt}\
\textbf{Multi modality}       &  \textbf{ -6.68 $\pm$ 0.31} & \textbf{181.5 $\pm$ 25.88} \\  
CT modality       & -6.69 $\pm$ 0.28  & 184.16 $\pm$ 22.84 \\ 
Demographics + Lung Volume & -6.75 $\pm$ 0.33 & 185.52 $\pm$ 22.89    \\ \specialrule{1pt}{.5pt}{.5pt}
\end{tabular}}
\vspace{-10pt}
\end{table}
\noindent \textbf{Implementation Details.} 
We used publicly available PyTorch \cite{NEURIPS2019_9015} framework to implement our proposed Fibro-CoSANet and an Intel(R) Xeon(R) Gold 5118 CPU with 187 GB physical ram and an Nvidia Tesla V100 SXM2 (32GB) GPU to run our experiments. We trained models for 40 epochs using Adam optimizer with a decoupled weight decay regularization of 0.01. We initialized the backbone CNN by the ImageNet pre-trained model and optimized the network to minimize the $L1$ loss.

\vspace{-5pt} 
\subsection{Results of Proposed Fibro-CoSANet} 
We first conducted an ablation study to analyze the effectiveness of the multi-modality feature fusion technique by comparing it with other available modes. To demonstrate the superiority of our proposed \textit{Multi-modal} training pipeline, we conducted experiments under three different \textit{modes}: (i) \textbf{Multi modality}: convolutional features from CT images + shallow features (demographics + lung volume), (ii) \textbf{CT modality}: convolutional features from only CT modality, (iii) \textbf{Shallow Modality}: only lung volume and demographic features were used to train our model without any CNN backbone. We found that the multi-modality modes achieved higher performance than standalone CT modality and shallow modality in terms of $LLL_m$ and $RMSE$ (Table~\ref{tab:modality}). These results suggested that demographics with lung volume or CT scans independently achieve reasonable performance while combing these two modalities improved the overall performance. Note that, the reported experimental results in the following sections are based on only multi modalities.

%%%%%%%%%%%%%%%%%
\begin{table}
\centering
\scriptsize
\caption{Performance comparison of Fibro-CoSANet with recent works in terms of $ LLL_{m}$ and $RMSE$. Our proposed Fibro-CoSANet outperforms the existing state-of-the-art works on predicting the progression of pulmonary fibrosis.}

\label{tab:comparison_m}
\def\arraystretch{1.2}
\setlength\tabcolsep{3.9pt}
\resizebox{0.70\textwidth}{!}{
\begin{tabular}{c|l|c|c} 
	\specialrule{1.2pt}{1pt}{1pt}\
Work & Regression Type & $ LLL_{m} \downarrow$ & $RMSE \downarrow$ \\ 
\specialrule{1.2pt}{1pt}{1pt}\

Fibrosis-Net~\cite{wong_fibrosis-net_2021}& Elastic Net & --6.82  & -\\   
\midrule
\multirow{3}{*} {Mandal et al. \cite{mandal_prediction_2020}}  & Quantile  & --6.92  & - \\ 
 & Ridge  & --6.81 & - \\ 
 & Elastic Net   & --6.72  & - \\ \hline
\multirow{4}{*}{Fibro-CoSANet (\textbf{Ours}) } & EfficientNet-b2 & \textbf{--6.68 $\pm$ 0.31} & \textbf{181.5 $\pm$ 25.88}  \\ 
 & ResNet-50 & \textbf{--6.68 $\pm$ 0.31} & \textbf{181.6 $\pm$ 22.89} \\ 
 & EfficientNet-b3 &  \textbf{--6.68 $\pm$ 0.28} & 182.58 $\pm$ 24.04  \\ 
 & EfficientNet-b1  & \textbf{--6.68 $\pm$ 0.28} & 183.96 $\pm$ 22.89 \\

 \specialrule{1.2pt}{1pt}{1pt}

\end{tabular}}
% \vspace{-10pt}
\end{table}
\vspace{-5pt}
\subsection{Comparison with Recent Approaches} We compared the overall performance of our proposed method with recent state-of-the-art approaches which predict the progression of pulmonary fibrosis (Table~\ref{tab:comparison_m}). Mandal et al. used \textit{Multiple Quantile Regression}, \textit{Ridge Regression}, and \textit{Elastic Net Regression} to predict the progression, while Elastic Net Regression achieved the best result, achieving $-LLL_m$ of -6.72 \cite{mandal_prediction_2020}. Wong et al. achieved $-LLL_m$ of -6.82 \cite{wong_fibrosis-net_2021}. Our proposed algorithm with \textit{EfficientNet-b2} performed better than \cite{mandal_prediction_2020, wong_fibrosis-net_2021}, resulting in $-LLLm$ of --6.68 and $RMSE$ of $  181\pm 25.88$ (Table~\ref{tab:comparison_m}). Also, the approximate complexity of our model was linear with respect to the number of patients as we processed all the information of a patient in a single mini-batch.

%%%%%%%%%%%%%%%%
\begin{table}[t]
\centering
\scriptsize
\caption {Fibro-CoSANet results under different CNN backbone. \tablefootnote{$LLL_m$ = Modified Laplace Log Likelihood, RMSE = Root mean square error, CV = Cross valitaion (5-fold)}}   
\label{tab:baseline_results}
\def\arraystretch{1.25}
\setlength\tabcolsep{5.9pt}
\resizebox{0.60\textwidth}{!}{
\begin{tabular}{l|c|c}
	\specialrule{1.2pt}{1pt}{1pt}\
       Backbone & $LLL_{m} \downarrow $(CV) & $RMSE \downarrow$ (CV) \\ \specialrule{1.2pt}{1pt}{1pt}\
              ResNet-18       &   -6.70 $\pm$ 0.29      &    183.68 $\pm$ 23.52               \\ 
              ResNet-34       &   -6.72 $\pm$ 0.28     &     185.18 $\pm$ 22.71                  \\ 
              ResNet-50       &   -6.72  $\pm$ 0.27      &   186.52 $\pm$  21.03                    \\ 
              ResNet-101      &  -6.71 $\pm$ 0.25      &   188.92 $\pm$ 19.94                   \\ 
              ResNet-152      &   -6.73 $\pm$ 0.28     &   186.19 $\pm$ 21.75                      \\ 
              ResNeXt-50      &   -6.72 $\pm$ 0.27      &   186.39 $\pm$ 24.64                      \\ 
              ResNeXt-101     &   -6.70 $\pm$ 0.26      &   184.04 $\pm$ 22.62                      \\  \cline{1-3} 
  EfficientNet-b0 &   -6.70 $\pm$ 0.29  &   183.00 $\pm$ 23.60                  \\ 
              EfficientNet-b1 &   -6.72 $\pm$ 0.31  &   183.22 $\pm$ 23.35                 \\ 
              EfficientNet-b2 &  -6.74 $\pm$ 0.34  &   184.17 $\pm$ 22.89                                                    \\ 
              EfficientNet-b3 &  -6.74 $\pm$ 0.34  &   184.17 $\pm$ 22.89                 \\ 
              EfficientNet-b4 &  -6.70 $\pm$ 0.30  &  183.00 $\pm$ 22.42                \\ 
            \specialrule{1.2pt}{1pt}{1pt}
  \end{tabular}}
 \vspace{-10pt}
\end{table}

\begin{table}
\centering
\tiny
\caption{Performance of Fibro-CoSANet under different backbone with respect to attention module hyper-parameters. FS and SF refer to filter size and stacking factor, respectively. It is clear that stacking the self-attention module improves the overall performance.}
\label{tab:attention_result}
\def\arraystretch{1.18}
\setlength\tabcolsep{3.7pt}
\resizebox{0.42\textwidth}{!}{
\begin{tabular}{c|c|c|c|c}
	\specialrule{1.2pt}{1pt}{1pt}\
Backbone &FS &  SF & $LLL_m \downarrow $& $RMSE \downarrow$\\ 	\specialrule{1pt}{1pt}{1pt}\

\multirow{5}{*}{EfficientNet-B0} & 32 & 3 & -6.7 $\pm$ 0.29 & 183.7 $\pm$ 23.55 \\ 
 &32 & 5 & -6.77 $\pm$ 0.31 & 185.63 $\pm$ 21.33 \\
 & 64 & 1 & -6.72 $\pm$ 0.34 & 182.13 $\pm$ 22.63 \\
 & 128 & 3 & -6.73 $\pm$ 0.33 & 183.67 $\pm$ 24.56 \\
 & 128 & 5 & -6.74 $\pm$ 0.36 & 183.57 $\pm$ 22.55 \\
\hline
 
\multirow{5}{*}{EfficientNet-B1} & 32 & 3 & \textbf{-6.68 $\pm$ 0.28} & 183.96 $\pm$ 22.89 \\
 &32 & 5 & -6.7 $\pm$ 0.28 & 185.64 $\pm$ 24.25 \\
 & 64 & 1 & -6.71 $\pm$ 0.29 & 184.16 $\pm$ 24.78 \\
 &128 & 3 & -6.79 $\pm$ 0.38 & 18-6.31 $\pm$ 24.79 \\
 &128 & 5 & -6.69 $\pm$ 0.31 & 183.12 $\pm$ 22.05 \\
 \hline

\multirow{5}{*}{EfficientNet-B2} & 32 & 3 & \textbf{-6.68 $\pm$ 0.31} & \textbf{181.5 $\pm$ 25.88} \\
 & 32 & 5 & -6.69 $\pm$ 0.3 & 183.39 $\pm$ 21.98 \\
 & 64 & 1 & -6.73 $\pm$ 0.28 & 184.71 $\pm$ 20.74 \\
 & 128 & 3 & -6.77 $\pm$ 0.33 & 187.13 $\pm$ 21.03 \\
 & 128 & 5 & -6.75 $\pm$ 0.33 & 18-6.03 $\pm$ 23.14 \\
\hline

\multirow{5}{*}{EfficientNet-B3} & 32 & 3 & -6.72 $\pm$ 0.34 & 183.28 $\pm$ 22.87 \\
 & 32 & 5 & -6.74 $\pm$ 0.31 & 184.68 $\pm$ 21.05 \\
 & 64 & 1 & -6.71 $\pm$ 0.34 & 183.34 $\pm$ 22.57 \\
 & 128 & 3 & \textbf{-6.68 $\pm$ 0.28} & 182.58 $\pm$ 24.04 \\
 & 128 & 5 & -6.72 $\pm$ 0.33 & 184.01 $\pm$ 24.4 \\
 \hline
 
\multirow{5}{*}{EfficientNet-B4} & 32 & 3 & -6.73 $\pm$ 0.3 & 184.86 $\pm$ 22.6 \\

 & 32 & 5 & \textbf{-6.68 $\pm$ 0.3} & 183.45 $\pm$ 23.19 \\
 & 64 & 1 & -6.73 $\pm$ 0.39 & 182.29 $\pm$ 23.11 \\
 & 128 & 3 & -6.74 $\pm$ 0.32 & 184.35 $\pm$ 22.04 \\
 & 128 & 5 & -6.71 $\pm$ 0.28 & 184.06 $\pm$ 23.57 \\
 \hline
 
\multirow{5}{*}{ResNet-18}  & 32 & 3 & -6.73 $\pm$ 0.32 & 184.71 $\pm$ 23.79 \\ 
 & 32 & 5 & -6.71 $\pm$ 0.31 & 183.79 $\pm$ 21.39 \\
 & 64 & 1 & -6.69 $\pm$ 0.28 & 183.84 $\pm$ 21.9 \\
 & 128 & 3 & -6.72 $\pm$ 0.27 & 184.96 $\pm$ 22.54 \\
 & 128 & 5 & -6.73 $\pm$ 0.35 & 185.29 $\pm$ 24.12 \\
 \hline
 
\multirow{5}{*}{ResNet-34}  & 32 & 3 & -6.73 $\pm$ 0.31 & 184.79 $\pm$ 21.45 \\ 
 & 32 & 5 & -6.72 $\pm$ 0.28 & 185.4 $\pm$ 21.98 \\ 
 & 64 & 1 & -6.71 $\pm$ 0.31 & 183.3 $\pm$ 22.79 \\ 
 & 128 & 3 & -6.7 $\pm$ 0.29 & 183.32 $\pm$ 23.83 \\
 & 128 & 5 & -6.72 $\pm$ 0.28 & 184.33 $\pm$ 21.54 \\
 \hline
 
\multirow{5}{*}{ResNet-50}  & 32 & 3 & -6.72 $\pm$ 0.31 & 184.07 $\pm$ 21.74 \\
 & 32 & 5 & -6.7 $\pm$ 0.28 & 184.94 $\pm$ 22.52 \\
 & 64 & 1 & -6.73 $\pm$ 0.27 & 185.46 $\pm$ 22.6 \\
 & 128 & 3 & \textbf{-6.68 $\pm$ 0.31} & \textbf{181.6 $\pm$ 22.89} \\
 & 128 & 5 & -6.74 $\pm$ 0.34 & 185.06 $\pm$ 24.97 \\
 
 \hline

\multirow{5}{*}{ResNet-101}  & 32 & 3 & -6.71 $\pm$ 0.3 & 183.13 $\pm$ 24.87 \\
 & 32 & 5 & -6.69 $\pm$ 0.28 & 184.17 $\pm$ 23.38 \\
 & 64 & 1 & -6.73 $\pm$ 0.28 & 185.33 $\pm$ 23.05 \\
 & 128 & 3 & -6.72 $\pm$ 0.3 & 185.53 $\pm$ 22.72 \\
 & 128 & 5 & -6.7 $\pm$ 0.3 & 183.63 $\pm$ 22.39 \\\hline

\multirow{5}{*}{ResNet-152}  & 32 & 3 & -6.7 $\pm$ 0.29 & 183.89 $\pm$ 22.25 \\
 & 32 & 5 & -6.7 $\pm$ 0.3 & 183.64 $\pm$ 23.6 \\
 & 64 & 1 & -6.72 $\pm$ 0.33 & 183.05 $\pm$ 22.54 \\
 & 128 & 3 & -6.73 $\pm$ 0.35 & 184.15 $\pm$ 23.15 \\
 & 128 & 5 & -6.72 $\pm$ 0.33 & 182.92 $\pm$ 23.9 \\\hline

\multirow{5}{*}{ResNeXt-50}  & 32 & 3 & -6.7 $\pm$ 0.27 & 184.75 $\pm$ 22.75 \\
 & 32 & 5 & -6.71 $\pm$ 0.3 & 184.61 $\pm$ 23.26 \\
 & 64 & 1 & -6.71 $\pm$ 0.32 & 183.84 $\pm$ 20.58 \\
 & 128 & 3 & -6.73 $\pm$ 0.28 & 185.22 $\pm$ 22.57 \\
 & 128 & 5 & -6.73 $\pm$ 0.32 & 184 $\pm$ 22.88 \\\hline

 \multirow{5}{*}{ResNeXt-101} & 32 & 3 & -6.72 $\pm$ 0.28 & 184.01 $\pm$ 23.38 \\
  & 32 & 5 & -6.7 $\pm$ 0.3 & 183.46 $\pm$ 23.37 \\
  & 64 & 1 & -6.7 $\pm$ 0.31 & 182.5 $\pm$ 24.82 \\
  & 128 & 3 & -6.72 $\pm$ 0.3 & 184.9 $\pm$ 22.38 \\
  & 128 & 5 & -6.7 $\pm$ 0.28 & 183.57 $\pm$ 22.21 \\ 	\specialrule{1.2pt}{1pt}{1pt}
\end{tabular}}
\vspace{-10pt}
\end{table}

%% file: discussion.tex
\vspace{-5pt}
\subsection{Baseline Analysis of Fibro-CoSANet}

We conducted an extensive experimental evaluation using widely-used CNNs, including \textit{ResNet}, \textit{ResNeXt}, and \textit{EfficientNet} to show the consistency and generalizability of our proposed approach. We reported experimental results under two key variants of our proposed pipeline as follows: \\
\noindent \textbf{Baseline Model without Self-Attention Module.} We implemented the base model under different network backbones without any self-attention layer. To show the consistency and  generalizability of our approach, we used  $12$ different CNNs architectures (five of ResNets, two of ResNeXts and five of EfficientNets) as the feature extractor for our proposed Fibro-CoSANet. Table~\ref{tab:baseline_results} presents the baseline results in terms of $LLL_m$  and $RMSE$. Interestingly, ResNets with lighter architecture (e.g., \textit{ResNet-18}- $LLL_m$: -6.70  RMSE: 183.68) and \textit{ResNeXt-101} achieved lower $LLL_m$ and $RMSE$ compared to the deeper ResNets. Furthermore, \textit{EfficientNet-b0}, and \textit{EfficientNet-b4} achieved comparative performance ($LLL_m \approx-6.70$ and $RMSE \approx 183.00$) to \textit{ResNet-18} . These results altogether suggested that our proposed Fibro-CoSANet with various backbones had the ability to predict FVC slope. The heavier models were prone to over-fitting as the size of the dataset was relatively smaller.
%%%%%%%%%%%%%%%%%%%%%%%%%
\begin{table}
\centering

\caption{Comparison of different variants of our proposed Fibro-CoSANet in terms of Macs (G), parameters (M), inference time (s), $LLL_m$, and $RMSE$.}

\def\arraystretch{1.25}
\setlength\tabcolsep{3.9pt}
\resizebox{0.65\textwidth}{!}{
\begin{tabular}{l|c|c|c|c|c}
\specialrule{1.2pt}{1pt}{1pt}\
Backbone & Macs (G) & Params (M) & Infer & $LLL_m \downarrow$ & $RMSE \downarrow$ \\ \specialrule{1.2pt}{1pt}{1pt}

EfficientNet-B0               & \textbf{0.07}        & \textbf{4.05}       & \textbf{0.67}  & -6.7 $\pm$ 0.29 & 183.7 $\pm$ 23.55\\ 
EfficientNet-B1               & 0.1         & 6.65       & 0.7 &\textbf{-6.68 $\pm$ 0.28} & 183.96 $\pm$ 22.89  \\ 
EfficientNet-B2               & 0.1         & 7.75       & 0.73 &\textbf{-6.68 $\pm$ 0.31} & \textbf{181.5 $\pm$ 25.88} \\ 
EfficientNet-B3               & 0.14        & 10.75      & 0.7  & -6.72 $\pm$ 0.34 & 183.28 $\pm$ 22.87 \\ 
EfficientNet-B4               & 0.18        & 17.61      & 0.71 &-6.73 $\pm$ 0.3 & 184.86 $\pm$ 22.6 \\ \hline
ResNet-18                      & 9.11        & 11.19      & 0.67 &-6.73 $\pm$ 0.32 & 184.71 $\pm$ 23.79 \\ 
ResNet-34                      & 18.79       & 21.3       & 0.63  &-6.73 $\pm$ 0.31 & 184.79 $\pm$ 21.45\\ 
ResNet-50                      & 21.13       & 23.57      & 0.69  &-6.7 $\pm$ 0.27 & 184.75 $\pm$ 22.75\\ 
ResNet-101                     & 40.61       & 42.56      & 0.69  &-6.71 $\pm$ 0.3 & 183.13 $\pm$ 24.87\\ 
ResNet-152                     & 60.1        & 58.21      & 0.7   & -6.7 $\pm$ 0.29 & 183.89 $\pm$ 22.25\\ 
ResNeXt-50                     & 21.92       & 23.04      & 0.68 &-6.7 $\pm$ 0.27 & 184.75 $\pm$ 22.75 \\ 
ResNeXt-101                    & 85.84       & 8-6.81      & 0.7  &-6.72 $\pm$ 0.28 & 184.01 $\pm$ 23.38 \\ \specialrule{1.2pt}{1pt}{1pt}

\end{tabular}}
\label{tab:complexity_n}
\vspace{-5pt}
\end{table}

\noindent \textbf{Baseline Model with Self-Attention Module.} 
To further improve the overall performance, we introduced a stacked self-attention module (Sec. \ref{selfattention}) on top of the each CNN backbone (Table~\ref{tab:attention_result}). Here, we used fixed output \textit{channel} dimension (32) of CNN backbones with several attention filter sizes, such as 32, 64, and 128 along with a different number of stacking factors, such as 1, 3, and 5 to empirically identify the best combination that achieved superior performance. As shown in Table \ref{tab:attention_result}, the overall performance was improved for most of the models with the addition of the self-attention layer. For instance, \textit{EfficientNet}-B1, B2, B3, B4, and \textit{ResNet-50}  improved the overall performance by a considerable margin, resulting in $\approx -6.68$ $LLL_m$. \textit{EfficientNet}-B2 and \textit{RestNet-50} achieved better score than other models in terms of $RMSE$($\approx 181$). Comparing the results of different variants under various design choices, \textit{EfficientNet-B2} achieved overall best performance ($ LLL_m : -6.68 \pm 0.31$ and $ RMSE: 181.5 \pm 25.88$) followed by \textit{RestNet-50} ($ LLL_m: -6.68 \pm 0.31$ and $RMSE : 181.6 \pm 22.89$) compared to other variants. We empirically found that \textit{EfficientNet-B2} and \textit{Reset-50} achieved the best performance with the attention filter size of 32 and 128, respectively, and three attention layers. ResNeXt-101 results were further not improved with the addition of self-attention later.

\vspace{-5pt}

\subsection{Performance Analysis} 

We analyzed the overall performance of our proposed approach under two key aspects: (i) Efficiency and (ii) Computational complexity.  \\
\noindent \textbf{Efficiency.}
One of the important aspects of the high-volume biomedical data analysis is the latency or inference speed of the system. Our approach used a single CT image and shallow modality features to calculate the prognosis line from a single scalar, $a$. This simple \textit{linear priori} assumption made the training and inference much faster, making our pipeline much efficient in handling a large amount of data. Note that, that the training complexity depends on the number of patients. \\
\noindent \textbf{Computational complexity of CNNs.} 
Table~\ref{tab:complexity_n} presents the comparison results of different baselines models in terms of the total number of parameters, inference time, and memory. Note that, we reported the best result for each CNN used in our experiments. \textit{EfficientNet-B0} achieved the lowest computational complexity (0.07  GMacs, 4.05 million parameters, 0.67 s inference) compared to other CNNs backbones; however, failed to achieve superior performance. This could be due to the fact that the \textit{EfficientNet-B0} architectures were relatively light-weight compared to ResNets. As \textit{EfficientNet-B2} achieved the best result with relatively lower computational complexity (0.1  GMacs, 7.75 million parameters, 0.73 s inference), we termed \textit{EfficientNet-B2} as the best network for FVC slope prediction. 

\vspace{-5pt}
\section{Discussion and Conclusion}
We proposed a novel multi-modal convolutional self-attention-based learning pipeline to predict the prognosis of IPF. To the best of our knowledge, this work was one of the earliest attempts that incorporated both CT scan and demographic information in an end-to-end manner. Furthermore, we integrated a self-attention layer on top of the CNNs to further refine the convolutional features by allowing the network to focus on a specific region of the CT scan image. Moreover, we predicted the slope of the FVC trend of a patient based on a simple linear prior assumption. Extensive experiments demonstrated the superiority of our proposed approach over the recent models tested on the same dataset \cite{mandal_prediction_2020,wong_fibrosis-net_2021}. 

Despite the impressive performance, one major limitation of our proposed approach was that the prognosis of pulmonary fibrosis was of linear nature. This assumption limited us to predict the actual FVC values at each temporal point. However, it allowed us to solve the problem as a simple regression problem for predicting the future prognosis status. Fibro-CoSANet failed to produce better performance with deeper architectures, which could be due to the relatively small sample size. Finally, throughout our experiments, we used a fixed set of hyper-parameters. The overall performance for each backbone can be further improved by a careful selection of the best possible hyper-parameters.

In conclusion, we aimed to provide a framework to the research community that can be used on a larger dataset and clinical trial in the future. As accurate progression prediction of IPF patients is crucial for the effective treatment and IPF based datasets are rarely available, our proposed algorithm could shed light on the new approaches to build trustworthy algorithms for IPF prognosis.